\documentclass[prd,%twocolumn,
nofootinbib,aps,longbibliography,showpacs]{revtex4-2}
\usepackage{amsmath,amssymb,mathtools}
\usepackage{mathrsfs}
\usepackage{tensor}
\usepackage{color,xcolor}
\usepackage{hyperref}

\hypersetup{colorlinks=true,
            linkcolor=magenta,
            filecolor=black,
            urlcolor=magenta,
            citecolor=magenta
           }

\allowdisplaybreaks
\begin{document}
%%%%%%%%%%%%%%%%%%%%%%%%%%%%%%%%%%%%%%%%%%%%%%%%%%%%%%%%%%%%%%%%%%%%%%%%%%%%
\title{Reconstruction of Black Hole Metric with Gravitational Shadows}

\author{Xing-hua Jin} 
%\email{jinxh@sbs.edu.cn}
\affiliation{Department of Mathematics, Shanghai Business School, Shanghai 200235, China}

\author{Dao-Jun Liu}
\email{djliu@shnu.edu.cn}
\affiliation{Department of Physics, College of mathematics and science, Shanghai Normal University, Shanghai 200234, China}

\date{\today}

%%%%%%%%%%%%%%%%%%%%%%%%%%%%%%%%%%%%%%%%%%%%%%%%%%%%%%%%%%%%%%%%%%%%%%%%%%%%
\begin{abstract}
Based on a two-parameter perturbative framework, we derive perturbative formulae for the radii of massive particle spheres and massive shadow radii, where an extremal black hole is employed as the perturbative background. Adopting extremal black holes as the perturbative background enables energy-dependent scans to probe stronger gravitational regimes.
Using these formulae, we perform perturbative analyses of the massive shadows for several non-standard black holes, and obtain the expansion expressions near the photon sphere of the extremal black hole. 
It is found that, although the dependence on the energy parameter \(\epsilon\) is the same across models, the coefficients associated with the deviation parameter \(\delta\) differ significantly, which can serve to distinguish among different black hole models. 
Furthermore, we construct a two‑point Padé approximant in the form of a continued fraction which achieves a globally accurate approximation for the massive shadow radius of the black hole over the entire parameter range.
Numerical tests show that the relative errors based on this approximant are small enough to provide a reliable foundation for model‑independent reconstruction of metric parameters.
\end{abstract}
%%%%%%%%%%%%%%%%%%%%%%%%%%%%%%%%%%%%%%%%%%%%%%%%%%%%%%%%%%%%%%%%%%%%%%%%%%%%


\maketitle
%%%%%%%%%%%%%%%%%%%%%%%%%%%%%%%%%%%%%%%%%%%%%%%%%%%%%%%%%%%%%%%%%%%%%%%%%%%%
\section{Introduction}\label{sec:intro}

Black hole shadows, as the most direct observational signals from the strong gravity regime, have become an important platform for testing gravitational theories in recent years due to the successful imaging of M87* and Sgr A* by the Event Horizon Telescope (EHT) \cite{EventHorizonTelescope:2019dse, EventHorizonTelescope:2022wkp, EventHorizonTelescope:2024dhe}. However, observations relying solely on photon shadows suffer from obvious theoretical degeneracy: many different metric configurations can produce almost the same shadow radius\cite{Cunha:2018acu, Lima:2021las, Vagnozzi:2022moj}. To overcome this limitation, a two-parameter perturbation theoretical framework has been recently developed\cite{Kobialko:2024zhc, Pantig:2025deu}, which systematically analyzes the corrections to the gravitational shadow radius from two aspects: particle energy and metric deviation.

Kobialko and Gal'tsov(\cite{Kobialko:2024zhc}, KG in short) constructed a perturbation theory for gravitational shadow in static spherically symmetric spacetimes and obtained a second-order expansion expression of the massive shadow radius with respect to the particle energy parameter $\epsilon$ and the metric deviation parameter $\delta$. 
They have pointed out that at the first-order approximation of the perturbation theory, the massive shadows of the Reissner-Nordstr\"om (RN) black hole, Bardeen black hole, Ghosh-Culetu-Simpson-Visser (GCSV) black hole, and magnetically charged Einstein-Euler-Heisenberg (EEH) black hole are indistinguishable, but at the second-order, all these models can be effectively distinguished. 
However, it is worth noting that KG's work assumed the area-radius gauge $\beta(r)=r^2$, i.e., only allowing the time component of the metric to be perturbed. 
To overcome this limitation, Pantig and Övgün\cite{Pantig:2025deu} generalized KG's work by allowing all metric functions to be perturbed simultaneously, thus applicable to a wider range of exotic compact objects, such as traversable wormholes and canonical scalar-tensor solutions. They explicitly proposed the concept of ``shadow spectroscopy'' for the first time, and demonstrated that different types of geometric deformations leave unique ``fingerprints'' in the energy dependence of massive shadow radii. This method provides a new theoretical tool for breaking degeneracies arising from the photon shadow alone and distinguishing standard black holes from exotic compact objects, laying a solid theoretical foundation for future multi-messenger observations.
{Refs.\cite{Kobialko:2024zhc, Pantig:2025deu} present expansions up to second order for both the massive particle sphere (MPS) radius and the massive shadow radius,
adopting the Schwarzschild black hole as the perturbation background.
But, further generalization to higher-order formulations requires rederivation owing to the implicit functional relations among \(r\), \(\epsilon\), and \(\delta\).
We herein provide an alternative derivation which can be readily extended to higher-order expressions. 
}

Furthermore, KG\cite{Kobialko:2024zhc} proposed that by observing massive shadows at multiple energies, one can formulate a system of linear equations for the ratio of the squared massive shadow radius to the squared photon shadow radius.
Solving this system allows the metric expansion coefficients to be determined independently, thereby achieving a model‑independent reconstruction of the metric parameters.
A notable advantage of this method is that it removes the dependence on both the black hole mass \(M\) and the observer distance, making the reconstruction scheme observationally more robust.
This means that even if there are some errors in the mass and distance determinations of M87* and Sgr A*, the conclusions of shadow spectroscopy reconstruction remain relatively reliable.
As illustrative examples, they discussed parameter reconstruction for the RN metric and $r^{-n}$ metric. 
It is found that high reconstruction accuracy is obtained for small parameter value, and the reconstruction error increases as the parameter grows larger, as will be demonstrated later.
To achieve globally high accuracy parameter reconstruction, we develop a two‑point Padé approximation method for the massive shadow radius.
It is well known that the two‑point Padé approximation\cite{BGM:96, Reboucas:2025maq, Reboucas:2026lbr} is designed to approximate a function whose power series expansions are known at two separate points.
The resulting approximant for the complete function naturally takes the form of a continued fraction.
Previous works \cite{Rezzolla:2014mua, Konoplya:2016jvv, Younsi:2016azx, Konoplya:2020hyk, Konoplya:2021slg, Mann:2021mnc, Konoplya:2022tvv, Maharana:2025lbm} based on Padé approximants mainly focused on constructing approximate metrics that satisfy the behavior in both the near-horizon and asymptotic regions.
Our approach expands the massive shadow radius not only around the Schwarzschild metric but also around the extremal black hole metric, and the two expansions are subsequently matched at their respective endpoints via a continued fraction ansatz.

The structure of this paper is as follows. 
In Sec.\ref{sec:massive}, we rederive the general perturbative formulae for the MPS radius and the massive shadow radius with respect to \(\epsilon\) and \(\delta\).
In Sec.\ref{PerEx}, we investigate the perturbative expansion of the MPS radius and the massive shadow radius the RN black hole around the extremal RN (ERN) black hole, and then apply this approach to several black hole models.
In Sec.\ref{sec:Reconstruction}, we reconstruct the black hole metric by means of Padé approximants.
In the end, a summary is given in Sec.\ref{sec:conclu}.
Besides, the complete expressions for the second-order partial derivatives are given in the Appendix \ref{sec:app}.

%%%%%%%%%%%%%%%%%%%%%%%%%%%%%%%%%%%%%%%%%%%%%%%%%%%%%%%%%%%%%%%%%%%%%%%%%%%%
\section{Perturbative Expansion of the Massive Shadows}\label{sec:massive}
	
In this section, we systematically introduces an alternative form of  perturbative expansion theory for the MPS radius and the gravitational shadow radius in static spherically symmetric spacetimes with respect to the particle energy parameter \(\epsilon\) and the metric deviation parameter \(\delta\). 
{Compared with existing works \cite{Kobialko:2024zhc, Pantig:2025deu}, we adopt the Implicit Function Theorem and the chain rule to present clearer partial derivative formulae, effectively avoiding the confusion in differentiation caused by implicit functional dependencies.}
	
For ease of comparison, we adopt the same notation for the static spherically symmetric metric as in Refs. \cite{Kobialko:2024zhc, Pantig:2025deu}:
\begin{equation}\label{metric}
ds^{2}=-\alpha(r,\delta)dt^{2}+\gamma(r,\delta)dr^{2}+\beta(r,\delta)\left(d\theta^{2}+\sin^{2}\theta\,d\phi^{2}\right),
\end{equation}
where \(\alpha, \beta, \gamma\) are smooth functions of the radial coordinate \(r\) and the deviation parameter \(\delta\). The parameter \(\delta\) describes the deviation of the metric relative to a reference background. 
Owing to the staticity and spherical symmetry, the system has two Killing vectors \(\xi_{(t)}^{\mu}=(1,0,0,0)\) and \(\xi_{(\phi)}^{\mu}=(0,0,0,1)\), corresponding respectively to the conserved energy \(E=-\xi_{(t)}^{\mu}p_{\mu}\) and angular momentum \(L=\xi_{(\phi)}^{\mu}p_{\mu}\). Without loss of generality, we restrict to the equatorial plane \(\theta=\pi/2\) when considering geodesic motion. The four-velocity \(u^{\mu}\) of a test particle with rest mass \(m\) satisfies the normalization condition \(g_{\mu\nu}u^{\mu}u^{\nu}=-m^{2}\). Using the conserved quantities, the radial equation of motion can be written as
\begin{equation}\label{ge}
\gamma(r,\delta)\dot{r}^{2}=\frac{E^{2}}{\alpha(r,\delta)}-\frac{L^{2}}{\beta(r,\delta)}-m^{2},
\end{equation}
where the dot denotes derivative with respect to an affine parameter. Introducing two dimensionless parameters \(\epsilon=m^{2}/E^{2}\) and \(l=L^{2}/E^{2}\), Eq. (\ref{ge}) can be rewritten as
\begin{equation}\label{ge1}
		\frac{\gamma(r,\delta)}{E^{2}}\dot{r}^{2}=V(r,\delta)\equiv \frac{1}{\alpha(r,\delta)}-\frac{l}{\beta(r,\delta)}-\epsilon,
\end{equation}
where \(V(r,\delta)\) is called the effective potential of the particle. The particle's orbit is determined by the shape of \(V(r,\delta)\).
	
The MPS corresponds to unstable circular orbits of the particle, determined by the extremum conditions of the effective potential
\begin{equation}\label{VdV}
		V(r,\delta)=0,\qquad V_{,r}(r,\delta)=0,
\end{equation}
where and hereafter the subscript comma before \(r\) denotes partial differentiation with respect to \(r\).
We can directly obtain the expressions of $\epsilon$ and $l$ from Eq. (\ref{ge1}) and condition (\ref{VdV}).
First, from \(V=0\) we obtain
\begin{equation}\label{epl}
	\epsilon =\dfrac{1}{\alpha(r, \delta)}\left[1-\frac{\beta(r, \delta)}{\alpha(r, \delta)}\frac{\alpha_{,r}(r, \delta)}{\beta_{,r}(r, \delta)}\right], 
\end{equation}
and substituting Eq. (\ref{epl}) into \(V_{,r}=0\) yields after simplification
\begin{equation}
	l=\frac{\beta(r, \delta)^{2}}{\alpha(r, \delta)^{2}}\frac{\alpha_{,r}(r, \delta)}{\beta_{,r}(r, \delta)}.
\end{equation}
Eq. (\ref{epl}) implicitly determines \(r\) as a function of \(\epsilon\) and \(\delta\), i.e., \(r=r(\epsilon,\delta)\). The MPS radius \(r_{\mathrm{MPS}}\) is defined as the solution of Eq. (\ref{epl}). Once the explicit metric functions are substituted into Eq. (\ref{epl}), we obtain the specific form of \(r_{\mathrm{MPS}}\), which can generally be expressed as a function of \(\epsilon\) and \(\delta\), \(r_{\mathrm{MPS}}=r_{\mathrm{MPS}}(\epsilon,\delta)\). Clearly, when \(\epsilon=0\), the MPS radius reduces to the usual photon sphere radius \(r_{\mathrm{PS}}\), namely \(r_{\mathrm{PS}}=r_{\mathrm{MPS}}(0,\delta)\). Unfortunately, only a few metrics, such as Schwarzschild and RN, admit analytic forms for \(r_{\mathrm{MPS}}\). For most metrics, no analytic expression exists, so a perturbative expansion is required.
	
We expand \(r_{\mathrm{MPS}}(\epsilon,\delta)\) around a reference point \((\epsilon_{0},\delta_{0})\) as follows
\begin{eqnarray}
	\nonumber
	&&r_{\mathrm{MPS}}(\epsilon,\delta)
=r_{\mathrm{MPS}}(\epsilon_0,\delta_0)
+r_{\mathrm{MPS},\epsilon}(\epsilon_0,\delta_0)
(\epsilon-\epsilon_0)
+r_{\mathrm{MPS},\delta}(\epsilon_0,\delta_0)(\delta-\delta_0)
\\\label{MPS}
&&\quad
+\dfrac{1}{2} r_{{\mathrm{MPS}},\epsilon\epsilon} {(\epsilon_0, \delta_0)} {(\epsilon - \epsilon_0)^2}
+ r_{{\mathrm{MPS}},\epsilon\delta}{(\epsilon_0, \delta_0)} (\epsilon - \epsilon_0)(\delta - \delta_0) 
+\dfrac{1}{2} r_{{\mathrm{MPS}},\delta\delta} 
{(\epsilon_0, \delta_0)} {(\delta - \delta_0)^2}
	+ \cdots.
\end{eqnarray}
When considering only the expansion in \(\epsilon\) , we obtain the expansion of \(r_{\mathrm{MPS}}\) around the photon sphere with \(\epsilon_{0}=0\)
\begin{equation}
	r_{{\mathrm{MPS}}}(\epsilon,\delta)
	=r_{{\mathrm{MPS}}}(0,\delta)
	+r_{{\mathrm{MPS}},\epsilon}(0,\delta)\epsilon
	+\frac{1}{2} r_{{\mathrm{MPS}},\epsilon\epsilon} {(0, \delta)} \epsilon^2 
	+\cdots.
\end{equation}
Similarly when considering only the expansion in \(\delta\) and setting \(\delta_{0}=0\), we obtain the expansion of \(r_{\mathrm{MPS}}\) around the base metric
\begin{equation}
r_{{\mathrm{MPS}}}(\epsilon,\delta)
=r_{{\mathrm{MPS}}}(\epsilon,0)
+r_{{\mathrm{MPS}},\delta}(\epsilon, 0)\delta
+\frac{1}{2} r_{{\mathrm{MPS}},\delta\delta} {(\epsilon, 0)} \delta^2 
	+\cdots.
\end{equation}

To compute the partial derivatives of \(r_{\mathrm{MPS}}\) with respect to \(\epsilon\) and \(\delta\) to various orders, we introduce an auxiliary function
\begin{equation}\label{F}
F(r, \epsilon, \delta) 
= \frac{1}{\alpha(r,\delta)}\left[1 - \frac{\beta(r,\delta)}{\alpha(r,\delta)} \frac{\alpha_{,r}(r,\delta)}{\beta_{,r}(r,\delta)}\right]
- \epsilon.
\end{equation}
It can be seen that \(F(r,\epsilon,\delta)\) has continuous partial derivatives in a neighborhood of the point \((r_{\mathrm{MPS}},\epsilon,\delta)\), and satisfies \(F(r_{\mathrm{MPS}},\epsilon,\delta)=0\) and \(F_{r}(r_{\mathrm{MPS}},\epsilon,\delta)\neq 0\). According to  the Implicit Function Theorem, the three-variable equation \(F(r,\epsilon,\delta)=0\) uniquely determines a continuous function \(r=r(\epsilon,\delta)\) with continuous partial derivatives in a neighborhood of that point, and we have
\begin{eqnarray}
	\label{dr1}
	&&r_{,\epsilon}=-\dfrac{F_{,\epsilon}}{F_{,r}},
	\\\label{dr2}
	&&
	r_{,\delta}=-\dfrac{F_{,\delta}}{F_{,r}}.
\end{eqnarray}
By further differentiating Eqs. (\ref{dr1}) and (\ref{dr2}), the second-order partial derivatives of \(r\) are obtained
\begin{eqnarray}\label{ddr1}
	&&	r_{,\epsilon\epsilon} = \dfrac{-F_{,\epsilon\epsilon} F_{,r}^2 + 2 F_{,\epsilon} F_{,r\epsilon} F_{,r} - F_{,\epsilon}^2 F_{,rr} }{F_{,r}^3},
	\\\label{ddr2}
	&&
	r_{,\delta\delta} = \dfrac{ -F_{,\delta\delta} F_{,r}^2 + 2 F_{,\delta} F_{,r\delta} F_{,r} - F_{,\delta}^2 F_{,rr} }{F_{,r}^3},
	\\\label{ddr3}
	&& r_{,\epsilon\delta} = \dfrac{ -F_{,\epsilon\delta} F_{,r}^2 + F_{,r} (F_{,\epsilon} F_{,r\delta} + F_{,\delta} F_{,\epsilon r}) - F_{,\epsilon} F_{,\delta} F_{,rr} }{F_{,r}^3}.
\end{eqnarray}
On the other hand, substituting Eq. (\ref{F}) into Eqs. (\ref{dr1}) and (\ref{dr2}), we obtain the explicit forms of the first-order partial derivatives:
\begin{eqnarray}
	\label{dr10}
	&&r_{,\epsilon}
	={\left(
		\dfrac{2 \beta  \alpha _{,r}^2}{\alpha ^3 \beta _{,r}}-\dfrac{2 \alpha _{,r}}{\alpha
			^2}+\dfrac{\beta  \alpha _{,r} \beta _{{,rr}}}{\alpha ^2 \beta
			_{,r}^2}-\dfrac{\beta  \alpha _{{,rr}}}{\alpha ^2 \beta _{,r}}
		\right)^{-1}
	},
	\\\label{dr20}
	&&
	r_{,\delta}=-
	\dfrac{
		\left(-\dfrac{\alpha _{,\delta }}{\alpha ^2}+\dfrac{2 \beta  \alpha _{,\delta } \alpha
			_{,r}}{\alpha ^3 \beta _{,r}}-\dfrac{\beta _{,\delta } \alpha _{,r}}{\alpha ^2 \beta
			_{,r}}-\dfrac{\beta  \alpha _{{,r\delta}}}{\alpha ^2 \beta
			_{,r}}+\dfrac{\beta  \alpha _{,r} \beta _{{,r\delta }}}{\alpha ^2 \beta
			_{,r}^2}\right)
	}
	{\left(
		\dfrac{2 \beta  \alpha _{,r}^2}{\alpha ^3 \beta _{,r}}-\dfrac{2 \alpha _{,r}}{\alpha
			^2}+\dfrac{\beta  \alpha _{,r} \beta _{{,rr}}}{\alpha ^2 \beta
			_{,r}^2}-\dfrac{\beta  \alpha _{{,rr}}}{\alpha ^2 \beta _{,r}}
		\right)
	}.
\end{eqnarray}
Similarly, substituting Eq. (\ref{F}) into Eqs. (\ref{ddr1})-(\ref{ddr3}) yields the explicit forms of the second-order partial derivatives of \(r\). Since these expressions are very complicated, we present them in Appendix \ref{sec:app}. Although these formulae are complex, they are easy to implement numerically and are applicable to any static spherically symmetric metric. Given specific metric functions \(\alpha\) and \(\beta\), once \(\epsilon_{0}\) and \(\delta_{0}\) are fixed, we can obtain the numerical values of the first-order and second-order partial derivatives of \(r_{\mathrm{MPS}}\) with respect to \(\epsilon\) and \(\delta\), and then substitute them into Eq. (\ref{MPS}) to obtain the approximate expression of the MPS radius near \((\epsilon_{0},\delta_{0})\).

In astrophysical observations, the observer is typically very far from the black hole, often at a  distance on the order of cosmological scale, so in practice one takes this distance to be infinity. By relating the properties of particles on their unstable circular orbits to the apparent angles on the celestial sphere of a distant observer, the squared gravitational shadow radius \(R^{2}(\epsilon,\delta)\) is obtained as \cite{Kobialko:2024zhc, Pantig:2025deu}
	\begin{equation}\label{R}
		R^{2}(\epsilon,\delta)=\left[\frac{\beta(r,\delta)}{\alpha(r,\delta)}\frac{1-\alpha(r,\delta)\epsilon}{1-\epsilon}\right]_{r_{\mathrm{MPS}}}.
	\end{equation}
	For the photon shadow (\(\epsilon=0\)), we recover the standard form \cite{Perlick:2021aok}
	\begin{equation}\label{RPS2}
		R_{\mathrm{PS}}^{2}=R^{2}(0,\delta)=\left[\frac{\beta(r,\delta)}{\alpha(r,\delta)}\right]_{r_{\mathrm{PS}}},
	\end{equation}
	where \(r_{\mathrm{PS}}\) is determined by the equation
	\begin{equation}\label{rps}
		\left[\frac{\alpha_{,r}(r,\delta)}{\alpha(r,\delta)}-\frac{\beta_{,r}(r,\delta)}{\beta(r,\delta)}\right]_{r_{\mathrm{PS}}}=0.
	\end{equation}
	
Similar to the expansion of the MPS radius, the squared gravitational shadow radius can be expanded as
\begin{eqnarray}
\nonumber
&&R^2(\epsilon,\delta)=R^2(\epsilon_0,\delta_0)+R^2_{,\epsilon}(\epsilon_0,\delta_0)(\epsilon-\epsilon_0)
+R^2_{,\delta}(\epsilon_0,\delta_0)(\delta-\delta_0)
\\\label{Rs}
&&\quad+\dfrac{1}{2}R^2_{,\epsilon\epsilon}(\epsilon_0,\delta_0){(\epsilon-\epsilon_0)^2}
+R^2_{,\epsilon\delta}(\epsilon_0,\delta_0)
(\epsilon-\epsilon_0)(\delta-\delta_0)
+\dfrac{1}{2}R^2_{,\delta\delta}(\epsilon_0,\delta_0)
{(\delta-\delta_0)^2}
	+\cdots
\end{eqnarray}
where \(R_{,\epsilon}^{2}\) and \(R_{,\delta}^{2}\) are the first-order partial derivatives of the squared gravitational shadow radius, and \(R_{,\epsilon\epsilon}^{2}, R_{,\delta\delta}^{2}, R_{,\epsilon\delta}^{2}\) are the second-order ones.
When considering only the expansion in \(\epsilon\) and taking \(\epsilon_{0}=0\), we have the expansion around the photon shadow
\begin{equation}
R^{2}(\epsilon,\delta)
=R^{2}(0,\delta)+R_{,\epsilon}^{2}(0,\delta)\epsilon
+\dfrac{1}{2}R_{,\epsilon\epsilon}^{2}(0,\delta){\epsilon^{2}}
+\cdots.
\end{equation}
When considering only the expansion in \(\delta\) and taking \(\delta_{0}=0\), we have the expansion around base metric
\begin{equation}
R^{2}(\epsilon,\delta)
=R^{2}(\epsilon, 0)+R_{,\delta}^{2}(\epsilon, 0)\delta
+\dfrac{1}{2}R_{,\delta\delta}^{2}(\epsilon, 0){\delta^{2}}
	+\cdots.
\end{equation}

To compute these partial derivatives more clearly, we define an auxiliary function
\begin{equation}\label{G}
		G(r,\epsilon,\delta)=\frac{\beta(r,\delta)}{\alpha(r,\delta)}\frac{1-\alpha(r,\delta)\epsilon}{1-\epsilon}.
\end{equation}
	Note that \(R^{2}(\epsilon,\delta)=G(r_{\mathrm{MPS}}(\epsilon,\delta),\epsilon,\delta)\). By the chain rule, the first-order partial derivatives of the squared gravitational shadow radius are
\begin{eqnarray}
	\label{R2e}
	&&R^2_{,\epsilon}
	=(G_{,\epsilon} +G_{,r} r_{,\epsilon})\big|_{r_{\text{MPS}}} ,
	\\\label{R2d}
	&&R^2_{,\delta}
	=(G_{,\delta} +G_{,r} r_{,\delta})\big|_{r_{\text{MPS}}} ,
\end{eqnarray}
where
\begin{eqnarray}
	\label{Ge}
	&&G_{,\epsilon}
	=\dfrac{\beta}{\alpha}
	\dfrac{(1-\alpha)}{(1-\epsilon)^2},
	\\\label{Gd}
	&&G_{,\delta}
	= - \dfrac{\beta}{\alpha(1 - \epsilon)}  
	\left[{ \dfrac{\alpha_{,\delta}}{\alpha} - 
		(1 - \alpha \epsilon) \dfrac{\beta_{,\delta}}{\beta} }
	\right],
	\\\label{Gr}
	&&G_{,r}
	= - \dfrac{\beta}{\alpha(1 - \epsilon)}  
	\left[{\dfrac{\alpha_{,r}}{\alpha}-(1-\alpha \epsilon) \dfrac{\beta_{,r}}{\beta} }
	\right].
\end{eqnarray}
Using relation (\ref{epl}), we find that \(G_{,r}|_{r_{\mathrm{MPS}}}=0\).  By substituting Eqs. (\ref{Ge})-(\ref{Gr}) into Eqs. (\ref{R2e}) and (\ref{R2d}), the first-order partial derivatives  are simplified to
\begin{eqnarray}\label{R1}
	&&R^2_{,\epsilon}
	=\left[ \frac{\beta}{\alpha} \frac{1-\alpha }{ (1-\epsilon)^2 }\right]_{r_{{\mathrm{MPS}}}},
	\\\label{R2}
	&&R^2_{,\delta}=\left\lbrace -\dfrac{\beta }{\alpha(1-\epsilon)}
	\left[ {\dfrac{\alpha_{,\delta} }{\alpha }-(1-\alpha  \epsilon )\dfrac{\beta_{,\delta}  }{\beta }}\right]\right\rbrace _{r_{{\mathrm{MPS}}}}.
\end{eqnarray}
Similarly, by the chain rule, the second-order partial derivatives of the squared gravitational shadow radius are
\begin{eqnarray}
	\label{R2ee}
	&&R^2_{,\epsilon\epsilon}
	=\left( G_{,\epsilon\epsilon}+2G_{,\epsilon r}r_{,\epsilon}
	+ G_{,r r}r_{,\epsilon}^2
	+ G_{,r} r_{,\epsilon\epsilon}\right)\big|_{r_{{\mathrm{MPS}}}} ,
	\\\label{R2dd}
	&&R^2_{\delta\delta}
	=\left( G_{,\delta\delta}+2G_{,\delta r}r_{,\delta}
	+ G_{,r r}r_{,\delta}^2
	+ G_{,r} r_{,\delta\delta}\right)\big|_{r_{{\mathrm{MPS}}}},
	\\\label{R2ed}
	&&R^2_{,\epsilon\delta} 
	=\left( G_{,\epsilon\delta}+G_{,\epsilon r}r_{,\delta}
	+ 
	G_{,r\delta}r_{,\epsilon}+G_{,r r}r_{,\delta}r_{,\epsilon}
	+
	G_{,r} r_{,\epsilon\delta}\right)\big|_{r_{{\mathrm{MPS}}}}.
\end{eqnarray}
The second-order derivatives of \(G\) can be simplified by using relation (\ref{epl}) into the following forms
\begin{eqnarray}
	\label{Gee}
	&&
	G_{,\epsilon\epsilon}\big|_{r_{{\mathrm{MPS}}}}
	=\left[ \frac{\beta}{\alpha} \frac{2(1-\alpha)}{(1-\epsilon)^3}\right]_{r_{{\mathrm{MPS}}}},
	\\\label{Gdd}
	&&G_{,\delta\delta}\big|_{r_{{\mathrm{MPS}}}}
	=\left[ \dfrac{\alpha  \beta_{,\delta\delta} (1-\alpha  \epsilon )-2 \alpha_{,\delta} \beta_{,\delta}
		-\beta  \left(  \alpha_{,\delta\delta}-2 \alpha^{-1}\alpha_{,\delta}^2\right)
		+\alpha\beta_{,\delta}}
	{\alpha^2 (1-\epsilon )}\right]_{r_{{\mathrm{MPS}}}},
	\\\label{Grr}
	&&G_{,r r}\big|_{r_{{\mathrm{MPS}}}}
	=\left[ \dfrac{\beta}{\alpha(1-\epsilon)}  
	\left(
	\dfrac{\beta_{,r}}{\beta}
	- 2\epsilon\dfrac{\alpha_{,r}\beta_{,r}}{\beta}
	+\dfrac{\alpha_{,r}\beta_{,rr}}  {\alpha\beta_{,r}}
	-\dfrac{\alpha_{,rr}}{\alpha}
	\right)\right]_{r_{{\mathrm{MPS}}}},
	\\\label{Ger}
	&&G_{,\epsilon r}\big|_{r_{{\mathrm{MPS}}}}
	=
	G_{,r\epsilon}\big|_{r_{{\mathrm{MPS}}}}
	=
	\left[ - \frac{\beta_{,r}}{(1-\epsilon)} \right]_{r_{{\mathrm{MPS}}}},
	\\\label{Gdr}
	&&G_{,\delta r}\big|_{r_{{\mathrm{MPS}}}}
	=
	G_{,r\delta}\big|_{r_{{\mathrm{MPS}}}}
	=
	\left\{ \dfrac{\beta}{\alpha(1-\epsilon)}  
	\left[ 
	\left(\dfrac{\beta_{,r}}{\beta}
	- \epsilon\dfrac{\alpha_{,\delta}\beta_{,r}}{\beta}\right)
	+\dfrac{\alpha_{,r}\beta_{,r\delta}}  {\alpha\beta_{,r}}
	-\dfrac{\alpha_{,r\delta}}{\alpha}
	+\dfrac{\alpha_{,r}}{\alpha}
	\left(\dfrac{\alpha_{,\delta}}{\alpha}
	-\dfrac{\beta_{,\delta}}{\beta}\right) 
	\right]
	\right\}_{r_{{\mathrm{MPS}}}},
	\\\label{Ged}
	&&G_{,\epsilon\delta}\big|_{r_{{\mathrm{MPS}}}}
	=G_{,\delta\epsilon}\big|_{r_{{\mathrm{MPS}}}}
	=\left[-\frac{1}{\alpha^2 (1-\epsilon)^2} 
	\left(\beta\alpha_{,\delta}-\alpha\beta_{,\delta}
	+\alpha^2\beta_{,\delta}\right)
	\right]_{r_{{\mathrm{MPS}}}}.
\end{eqnarray}
Substituting Eqs. (\ref{Gee})-(\ref{Ged}) into Eqs. (\ref{R2ee})-(\ref{R2ed}), we obtain the explicit forms of the second-order partial derivatives of the squared gravitational shadow radius.
	
The above perturbative expansions require that \(\epsilon\) and \(\delta\) are sufficiently small around the expansion point to ensure rapid convergence of the series. For photons, \(\epsilon=0\). For massive particles, if the rest energy is much smaller than the total energy, \(\epsilon\) is typically very small. 
For example, for neutrinos with energies of order GeV and rest energies \(\sim 0.1\) eV, \(\epsilon\sim 10^{-20}\), so the perturbative expansion is extremely effective. For effectively mass photons in plasma \cite{Perlick:2017fio, Perlick:2023znh, Bezdekova:2022gib, Briozzo:2022mgg, Kobialko:2023qzo, Jin:2020emq}, \(\epsilon\) is still a small quantity.
The parameter \(\delta\) describes the deviation of the metric from the background. It is also typically required to satisfy \(|\delta|\ll 1\). Therefore, first-order or second-order expansions are sufficient to describe most physically interesting cases.
	
Throughout the derivation, we have introduced two three-variable functions \(F(r,\epsilon,\delta)\) and \(G(r,\epsilon,\delta)\), and used the Implicit Function Theorem and the chain rule to clearly obtain the first-order and second-order partial derivatives of the MPS radius and the squared gravitational shadow radius. 
{This method effectively avoids the differentiation confusion that might arise from implicit dependencies among \(r,\epsilon\) and \(\delta\) as noted in Refs. \cite{Kobialko:2024zhc, Pantig:2025deu}. 
It is worth noting that,  although our formulae are different in form from those in \cite{Kobialko:2024zhc, Pantig:2025deu}, by applying them to the examples given in those papers, the same results are obtained, confirming consistency.}

%%%%%%%%%%%%%%%%%%%%%%%%%%%%%%%%%%%%%%%%%%%%%%%%%%%%%%%%%%%%%%%%%%%%%%%%%%%%
\section{Shadows Expansion Around the Extremal Black Holes}\label{PerEx}

The perturbative expansion formulae obtained above, can not only handle expansions based on the Schwarzschild metric, but also on other black hole metrics. In this section, we take the ERN black hole as a new expansion base, i.e., we set \(\epsilon_0=0,\delta_0=0\) corresponding to the photon sphere of the ERN black hole. 

Extremal black holes themselves are both interesting and critically important objects in strong-field gravity. They are a special class of black hole solutions in general relativity in which the charge (or angular momentum) reaches its maximum value, so that the inner and outer horizons coincide into a degenerate horizon, the surface gravity vanishes, and the Hawking temperature is strictly zero \cite{Anderson:2000pg, Liberati:2000sq}. The most typical examples are the ERN black hole ($|Q|=M$) and the extremal Kerr black hole ($|a|=M$). 
The extremal solutions of black holes provide an ideal platform for studying the microscopic origin of black hole entropy. 
One of the most famous is the pioneering work of Strominger and Vafa \cite{Strominger:1996sh}, who successfully explained the Bekenstein-Hawking entropy of extremal black holes using D-brane configurations. 
In addition, extremal black holes exhibit many unusual properties at both classical and semiclassical levels \cite{Garfinkle:2011pj, Aretakis:2011ha, Aretakis:2012bm, Good:2020nmz, Dalui:2026msw}, which have long posed challenges to theoretical physicists.

Choosing the ERN black hole as the base has the following advantages. 
(i) It is a typical black hole solution in general relativity with a degenerate horizon. Its photon sphere radius of the ERN \(r_{\mathrm{ERN}}=2M\) is significantly smaller than the Schwarzschild value of \(3M\), so the shadow is smaller, and its MPS radius also has a relatively simple algebraic structure. 
(ii) The ERN black hole lies on the boundary between black holes (\(Q<M\)) and naked singularities (\(Q>M\)), so the sign of the deviation parameter \(\delta=Q/M-1\) directly determines whether the spacetime contains a horizon. Therefore, the shadow radius is sensitive to \(\delta\), making it easy to distinguish different theoretical models. 
(iii) Many solutions beyond general relativity reduce to the ERN black hole in certain parameter limits, so expansions around this base are naturally applicable to these models.
	
In this section, we first study the perturbative expansion of the MPS radius and the massive shadow radius for the RN black hole around the ERN black hole as an example. 
Then, we apply the method to different black holes and compare their massive shadow ``fingerprints'' with those of the standard RN black hole.
	
\subsection{Shadow Expansion of RN Black Hole around the ERN Black Hole}
	
The metric of the ERN black hole is obtained by setting \(Q=M\) in the general RN metric, i.e.,
\begin{equation}\label{metricRN}
\alpha=\dfrac{1}{\gamma}=1-\frac{2M}{r} +\frac{M^2}{r^2}, \quad \beta=r^2.
\end{equation}
Here we use the metric form (\ref{metric}) with \(\delta=0\) in \(\alpha(r,\delta)\) corresponding to the ERN metric.
Substituting Eq. (\ref{metricRN}) into Eq. (\ref{epl}), we obtain the equation determining the MPS radius \(r_{\mathrm{MERN}}\) of the ERN black hole
\begin{equation}\label{ERNmps}
	\frac{r^2 (2 M-r)}{(M-r)^3}=\epsilon.
\end{equation}
Solving Eq. (\ref{ERNmps}) gives the exact analytic MPS radius
\begin{equation}\label{RNrE}
r_{\text{MERN}}
=M{\left\{
	\dfrac{
		%--
		\sqrt[3]{4} 
		\left[ \sqrt[3]{9 (5-3 \epsilon )
			\epsilon +3 \sqrt{3} \sqrt{(\epsilon -1)^2 \epsilon  (27
				\epsilon -32)}-16}\right]^2
		%--
		+2 \sqrt[3]{-2} (3 \epsilon-4)}
	{6 (\epsilon -1)\sqrt[3]{9 (5-3 \epsilon ) \epsilon +3\sqrt{3}\sqrt{(\epsilon -1)^2 \epsilon  (27 \epsilon-32)}-16}}
	+\dfrac{3 \epsilon-2}{3\epsilon -3}\right\}}.
\end{equation}
Substituting Eq.(\ref{RNrE}) into Eq.(\ref{R}), we obtain the squared massive shadow radius \(R_{\mathrm{MERN}}^2\) of the ERN black hole
\begin{equation}\label{RNR2}
R^2_{\mathrm{MERN}}
= \dfrac{r_{\mathrm{MERN}}^2}{1-\epsilon}
\left[\left(1-\frac{2M}{r_{\mathrm{MERN}}} +\frac{M^2}{r_{\mathrm{MERN}}^2}\right)^{-1}-\epsilon\right].
\end{equation}
Taking \(\epsilon=0\), i.e., the photon case, solving Eq. (\ref{ERNmps}) gives the photon sphere radius of the ERN black hole
	\begin{equation}\label{M2}
		r_{\mathrm{ERN}} = 2M.
	\end{equation}
Substituting Eq. (\ref{M2}) into Eq. (\ref{RPS2}) gives the squared photon shadow radius of the ERN black hole
	\begin{equation}
		R_{\mathrm{ERN}}^2 = 16M^2.
	\end{equation}
It is found that the photon shadow radius of the ERN black hole is smaller than that of the Schwarzschild black hole, indicating that charge significantly shrinks the shadow.
	
To study the massive shadow expansion of general RN black holes near the ERN black hole, we can rewrite the RN metric as
\begin{equation}\label{dM}
	\alpha(r,\delta) = 1 - \frac{2M}{r} + \frac{M^2(1+\delta)^2}{r^2}, \quad \beta = r^2, 
\end{equation}
where \(\delta=Q/M-1\) is the dimensionless small parameter measuring the deviation from extremality. When \(\delta<0\), i.e., \(Q<M\), the non-ERN black hole still has two horizons; when \(\delta=0\), i.e., \(Q=M\), the black hole becomes ERN with a single degenerate horizon; when \(\delta>0\), i.e., \(Q>M\), the horizon disappears and the spacetime becomes a naked singularity.
Substituting Eq. (\ref{dM}) into Eq. (\ref{epl}) gives the equation determining the MPS radius \(r_{\mathrm{MRN}}\) for RN black holes
\begin{equation}\label{RNmps}
	\epsilon = \frac{r^2\left[ r(r-3M) + 2M^2(1+\delta)^2 \right]}{\left[ r(r-2M) + M^2(1+\delta)^2 \right]^2}. 
\end{equation}
Using the partial derivative formulae derived in Sec.\ref{sec:massive}, 
we can compute the partial derivatives of \(F(r,\epsilon,\delta)\) and \(G(r,\epsilon,\delta)\) at the photon sphere point \((r_0=2M,\epsilon_0=0,\delta_0=0)\).
Then substituting them into Eqs. (\ref{MPS}) and (\ref{Rs}) and performing a series of algebraic operations, we obtain
the MPS radius \(r_{\mathrm{MRN}}\) and the squared massive shadow radius \(R_{\mathrm{MRN}}^2\) for RN black hole around the ERN photon sphere point \((\epsilon_0=0,\delta_0=0)\) 
\begin{equation}\label{rmpse}
r_{\mathrm{MRN}}	
=2M+\dfrac{M}{4} \epsilon-4M\delta
+\dfrac{M}{8} \epsilon^2-18M\delta^2
+\dots
=r_{\mathrm{ERN}}\left(1+\dfrac{1}{8} \epsilon-2\delta
+\dfrac{1}{16} \epsilon^2-9\delta^2
+\dots\right),
\end{equation}
and
\begin{eqnarray}
	\nonumber
	&&R^2_{\mathrm{MRN}}=16M^2+12M^2\epsilon-32M^2\delta
	+\frac{23}{2}M^2\epsilon^2
	-16M^2\epsilon\delta
	-80M^2{\delta^2}+\dots
	\\\label{R2mrn}
	&&\quad
	=R^2_{\mathrm{ERN}}
	\left( 1+\dfrac{3}{4}\epsilon-2\delta
	+\frac{23}{32}\epsilon^2
	-\epsilon\delta	-5{\delta^2}
	+\dots\right) .
\end{eqnarray}
It is worth noting the effect of the $\delta$ term on the massive shadow, where the first-order contribution plays the dominant role.
For $\delta<0$, the first-order correction $-32M^2\delta$ is positive and hence the massive shadow radius increases; conversely, for $\delta>0$, the correction is negative and the massive shadow radius decreases.
If we take $|\delta|=0.1$ as an illustrative case, the squared massive shadow radius is modified by $3.2M^2$, which amounts to a relative change of $20\%$.

Considering only the massive shadow expansion in $\delta$ over the ERN metric, we substitute Eq.(\ref{RNrE}) into the formulae in Sec.\ref{sec:massive} and obtain
\begin{equation}
\label{expan1-1}
R^2_{\text{MRN}}
=B_0+B_1 \delta+B_2 \delta^2+\cdots,
\end{equation}
where the coefficients $B_i$ are expressed in terms of the parameter $\epsilon$ as
\begin{equation}\label{beta012}
\begin{cases}
		B_0
		=\left(\dfrac{r_{\text{MERN}}^4}
		{-M^2+3 r_{\text{MERN}} M-r_{\text{MERN}}^2}\right) 
		=16M^2+12M^2\epsilon+\dfrac{23}{2}M^2\epsilon^2+\cdots,	&\\
		B_1
		=-\left( \dfrac{2 r_{\text{MERN}}^4 M}
		{M^3-4 r_{\text{MERN}} M^2
			+4r_{\text{MERN}}^2 M-r_{\text{MERN}}^3}\right) 
		=-32M^2-16M^2\epsilon-15M^2\epsilon^2+\cdots,	&\\
		B_2=-\dfrac{\left(4 M^4 r_{\text{MERN}}^4-5 M^3r_{\text{MERN}}^5
			+6 M^2 r_{\text{MERN}}^6-M r_{\text{MERN}}^7\right)}
		{(M-r_{\text{MERN}})^3 \left(4 M^3-13r_{\text{MERN}} M^2 
			+7 r_{\text{MERN}}^2 M-r_{\text{MERN}}^3\right)}
		=-80M^2-24M^2\epsilon-\dfrac{55}{2}M^2\epsilon^2+\cdots,	&\\
		\cdots .&\\
	\end{cases}
\end{equation}

To prepare for the subsequent two-point Padé approximation, the squared massive shadow radius further needs to be expanded about the Schwarzschild background for which
$Q/M=1+\delta$ is chosen as the parameter characterizing the deviation from the Schwarzschild metric.
Then, the MPS radius and the squared massive shadow radius for RN black hole around the Schwarzschild photon sphere are given by
\begin{equation}\label{rmSch}
r_{\mathrm{MRN}}	
=3M+\dfrac{M}{3} \epsilon+\dfrac{5M}{27}\epsilon^2
-\dfrac{2M}{3}(1+\delta)^2
+\dots
=r_{\mathrm{Sch}}\left[1+\dfrac{1}{9}\epsilon
+\dfrac{5}{81} \epsilon^2-\dfrac{2}{9}(1+\delta)^2
+\dots\right],
\end{equation}
and
\begin{eqnarray}
\label{R2Sch}
&&R^2_{\mathrm{MRN}}=27M^2+18M^2\epsilon
+17M^2\epsilon^2
-9M^2{(1+\delta)^2}+\dots
=R^2_{\mathrm{Sch}}
\left[1+\dfrac{2}{3}\epsilon+\frac{17}{27}\epsilon^2
-\dfrac{1}{3}{(1+\delta)^2}
+\dots\right],
\end{eqnarray}
where $r_{\mathrm{Sch}}=3M$ and $R^2_{\mathrm{Sch}}=27 M^2$
denote the photon sphere radius and the squared photon shadow radius for the Schwarzschild black hole, respectively.
It should be noted that although Eqs. (\ref{rmSch}) and (\ref{R2Sch}) appear to differ in form from Eqs. (76) and (77) in Ref. \cite{Kobialko:2024zhc}, this discrepancy arises from the different choice of deviation parameters, and in fact the two sets of equations are fundamentally identical.
Eqs. (\ref{R2mrn}) and (\ref{R2Sch}) demonstrate the difference between the two expansions of the massive shadow  in terms of two different bases.

Considering only the massive shadow expansion in $Q/M$ over the Schwarzschild metric, we first set $\delta=-1$ in Eq. (\ref{RNmps}) and obtain the analytic expression for the MPS radius of the Schwarzschild black hole
\begin{equation}\label{MSch}
r_{\text{MSch}}
=\frac{3 -4\epsilon+\sqrt{9-8\epsilon}}{2(1-\epsilon)}M.
\end{equation}
Then the massive shadow expansion over the Schwarzschild metric can be obtained as 
\begin{equation}
	\label{expan0-1}
R^2_{\text{MRN}}
=A_0+A_1 (1+\delta)+A_2 (1+\delta)^2+\cdots,
\end{equation}
where the coefficients $A_i$ are expressed in terms of the parameter $\epsilon$ as
\begin{equation}\label{alpha012}
\begin{cases}
		A_0
		=\left( \dfrac{r_{\text{MSch}}^3}{r_{\text{MSch}}-4 M}\right)
		=27M^2+18M^2\epsilon
		+17M^2\epsilon^2+\cdots,	&\\
		A_1=0,	&\\
		A_2
		=-\left( \dfrac{r_{\text{MSch}}^2 M}{4 M-r_{\text{MSch}}}\right) =-9 M^2-5 M^2 \epsilon
		-\dfrac{41 M^2 \epsilon ^2}{9}+\cdots,	&\\
		\cdots .&\\
	\end{cases}
\end{equation}
Eqs. (\ref{expan1-1}) and (\ref{expan0-1}) show the difference between the two expansions of the massive shadow in two different bases.

\subsection{Application to other black hole models}

Our method is applicable not only to RN black holes but also to other black holes.
Here we consider four typical black hole models, namely the EEH black hole\cite{Vagnozzi:2022moj}, the charged black hole in Kalb-Ramond (KR) gravity\cite{Gross:1984dd, Duan:2023gng, Yang:2025byw}, the Frolov black hole\cite{Frolov:2016pav, Tang:2026mem, Muniz:2025ugk} and the charged black hole solution in modified gravity(MOG)\cite{Nishonov:2025hxz}.
Their metrics take the form of deviations from the ERN metric and can be written in a unified expression as follows
\begin{equation}
	\alpha = 1-\dfrac{2M}{r} +\dfrac{M^2}{r^2}
	+c_1\delta+c_2\delta^2+\cdots.
\end{equation}
Table \ref{t0} lists the explicit expressions for these four models.

\begin{table}[h]
\centering
\begin{tabular}{|l|c|c|c|}
\toprule
Metric & $\delta$ & $c_{1}$ & $c_{2}$ \\
\hline
RN& $\dfrac{Q}{M} -1$ & $ \dfrac{2M^2}{r^2}$ & $\dfrac{M^2}{r^2}$ \\
EEH & $\dfrac{\mu}{M^2}$ & $-\dfrac{2}{5}\dfrac{M^6}{r^6}$ & $0$  \\
Charged KR & $\ell$ & $1+2\dfrac{M^2}{r^2}$ &$1+3\dfrac{M^2}{r^2}$\\
Frolov & $\ell$ & $-\dfrac{M^4 (M-2 r) (M+2 r)}{r^6}$ & $\dfrac{M^7 (M-2 r) \left(M^2+4 M r+4 r^2\right)}{r^{10}}$  \\
Charged MOG & $\alpha_{\text{M}}$ & $\dfrac{2 \left(M^2-M r\right)}{r^2}$ & $\dfrac{M^2}{r^2}$  \\
	\hline
\end{tabular}
\caption{Expansion coefficients of metric function $\alpha$ for five typical  black holes.\label{t0}}
\end{table}

To illustrate the distinguishability of the above models under the ERN base, we summarize the first-order and second-order expansion coefficients of the MPS radius and the squared massive shadow radius in Table \ref{t1}. 

\begin{table}[h]
\centering
\begin{tabular}{|l|c|c|c|c|c|c|c|c|c|c|}
\toprule
Metric & $a_{10}$ & $a_{01}$ & $a_{20}$ & $a_{11}$ & $a_{02}$ & $b_{10}$ & $b_{01}$ & $b_{20}$ & $b_{11}$ & $b_{02}$ \\
\hline
RN& $1/8$ & $-2$ & $1/16$ & $0$ & $-9$ & $3/4$ & $-2$ & $23/32$ & $-1$ & $-5$ \\
EEH & $1/8$ & $1/20$ & $1/16$ & $-3/160$ & $-3/200$ & $3/4$ & $1/40$ & $23/32$ & $0$ & $-7/1600$ \\
Charged KR & $1/8$ & $-4$ & $1/16$ & $0$ & $-21$ & $3/4$ & $-6$ & $23/32$ & $-4$ & $-3$ \\
Frolov & $1/8$ & $-11/8$ & $1/16$ & $13/64$ & $-839/128$ & $3/4$ & $-15/16$ & $23/32$ & $-1/4$ & $-167/64$ \\
Charged MOG & $1/8$ & $1$ & $1/16$ & $1/8$ & $0$ & $3/4$ & $2$ & $23/32$ & $3/2$ & $1$ \\
\hline
\end{tabular}
\caption{Perturbation coefficients of the MPS radius and the squared massive shadow radius.\label{t1}}
\end{table}
The table lists the coefficients in the expansions 
\begin{equation}
r_{\mathrm{MPS}} = r_{\mathrm{ERN}}(1 + a_{10}\epsilon + a_{01}\delta + a_{20}\epsilon^2 + a_{11}\epsilon\delta + a_{02}\delta^2 + \cdots)
\end{equation}
and
\begin{equation}
R^2 = R_{\mathrm{ERN}}^2 (1 + b_{10}\epsilon + b_{01}\delta + b_{20}\epsilon^2 + b_{11}\epsilon\delta + b_{02}\delta^2 + \cdots).
\end{equation}
From the table, the following conclusions can be drawn:
\begin{itemize}
\item \(\epsilon\) dependence: 
All models share identical values for the coefficients \(a_{10}\), \(a_{20}\), \(b_{10}\), and \(b_{20}\) respectively.
Consequently, the variation of the massive shadow against \(\epsilon\) alone is insufficient to discriminate among any of the models within the ERN framework. 
This degeneracy underscores the necessity of including \(\delta\) dependence and higher-order mixed terms.

\item \(\delta\) dependence: 
The essential distinctions among models reside in the \(\delta\) related coefficients. 
The sign of the first-order coefficient \(b_{01}\) directly determines whether an increase in \(\delta\) enlarges or shrinks the shadow. 
Specifically, a positive \(\delta\) decreases the massive shadow radius for the RN, charged KR, and Frolov black holes, but increases it for the EEH and charged MOG black holes.  
The large spread of \(b_{01}\) values, ranging from \(-6\) to \(+2\), thus offers the most direct observational evidence for model discrimination.
The coefficient \(b_{02}\) measures the quadratic deviation of the massive shadow radius from a linear response in \(\delta\). 
For the RN, EEH, charged KR, and Frolov cases, \(b_{02}<0\), implying an additional reduction of the shadow; for the charged MOG case, however, \(b_{02}>0\), resulting in an increase, which exhibits a clear opposite tendency.
Hence, provided that constraints on \(\delta\) are imposed by independent observables (e.g., quasinormal modes and gravitational waves), the full-order dependence of the shadow radius on \(\delta\) can serve as a valid theoretical criterion for discriminating between the above models.

\item \(\epsilon\delta\) dependence: The coefficient \(b_{11}\) exhibits significant variation across models, suggesting that joint measurements of the dependencies on \(\epsilon\) and \(\delta\) can further improve model discrimination through the second-order cross-term.
\end{itemize}

These distinct behaviors demonstrate the power of shadow spectroscopy.
Specifically, by measuring massive shadow radii at multiple energies \(\epsilon\) and combining these measurements with independent or joint constraints on the deviation parameter \(\delta\), one can distinguish among black hole models beyond general relativity.
Ultimately, this approach enables a model-independent reconstruction of the background geometry of compact objects, thereby breaking the degeneracy inherent in single photon shadow observations.

It is worth noting that the photon sphere radius of the ERN black hole, \(2M\), is significantly smaller than the Schwarzschild value of \(3M\).
This smaller radius permits a radial scan, implemented by tuning the energy parameter \(\epsilon\), which explores stronger gravitational regimes and consequently provides greater sensitivity to the geometric structure of spacetime.
Moreover, a number of modified gravity models, such as EEH and KR, naturally reduce to the ERN black hole in certain parameter limits. Consequently, the differences in their expansion coefficients relative to the ERN baseline directly reflect the physical nature of each correction term, without introducing any additional free parameters.
By contrast, choosing the Schwarzschild metric as the base yields extra model-dependent parameters in the expansion, thereby increasing the degeneracy among different models.

%--------------------

\section{Reconstruction of Black Hole Metric Using Padé Approximants}\label{sec:Reconstruction}
Based on the above introduced formulation, we can further develop a new approach for reconstructing the parameters of a black hole metric.
As an illustrative example, we consider the RN metric.

To this end, we first rewrite Eqs. (\ref{expan0-1}) and (\ref{expan1-1}) as follows
\begin{eqnarray}
	\label{expan0}
	&&R^2_{\text{MRN}}
	=A_0+A_1 x+A_2 x^2+\cdots
	\\\label{expan1}
	&&R^2_{\text{MRN}}
	=B_0+B_1 (x-1)+B_2 (x-1)^2+\cdots
\end{eqnarray}
where $x=Q/M$.
The coefficients $A_i$ and $B_i$, expressed in terms of the parameter $\epsilon$, are given in Eqs. (\ref{beta012}) and (\ref{alpha012}), respectively.
It is easy to find that Eqs. (\ref{expan0}) and (\ref{expan1}) are only valid in the vicinity of  $x=0$ and $x=1$, respectively.
To obtain a better global approximation to the exact solution, a two-point Padé approximant can be constructed for the squared massive shadow radius.
Conveniently, the squared massive shadow radius for the RN black hole can be parametrized by Padé approximant in the form of continued fractions
\begin{equation}
	\label{padeMRN}
	R^2_{\text{MRN}}=B_0+B_1 (x-1)
	+\dfrac{a_1}
	{1+\dfrac{a_2 x}
		{1+\dfrac{a_3 x}{1+\dfrac{a_4 x}{1+\cdots}}
		}
	} (x-1)^2,
\end{equation}
where $a_1, a_2, a_3, \cdots$ are dimensionless constants to be constrained.
Eq. (\ref{padeMRN}) are introduced to describe the squared massive shadow radius near the Schwarzschild metric (i.e., for $x\approx 0$), as well as at the ERN metric (i.e., for  $x\approx 1$).

\begin{figure}[ht]
	\centering
	\includegraphics[width=0.8\textwidth]{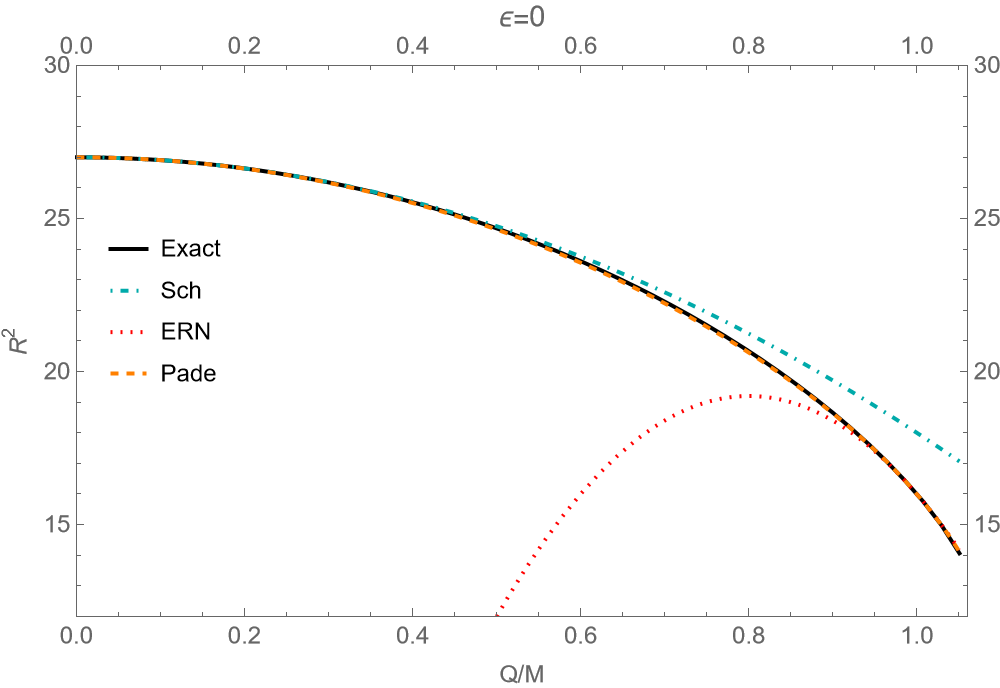}
	\caption{\small{The squared photon shadow radius $R^2$ of the RN black hole as a function of $Q/M$. The solid line represents the exact solution, while the dashed line corresponds to the Padé approximant. The dotdashed and dotted line denote the $\delta$-expansions over Schwarzschild metric and the ERN metric, respectively.}}\label{fig1}
\end{figure}

\begin{figure}[ht]
	\centering
	\includegraphics[width=0.8\textwidth]{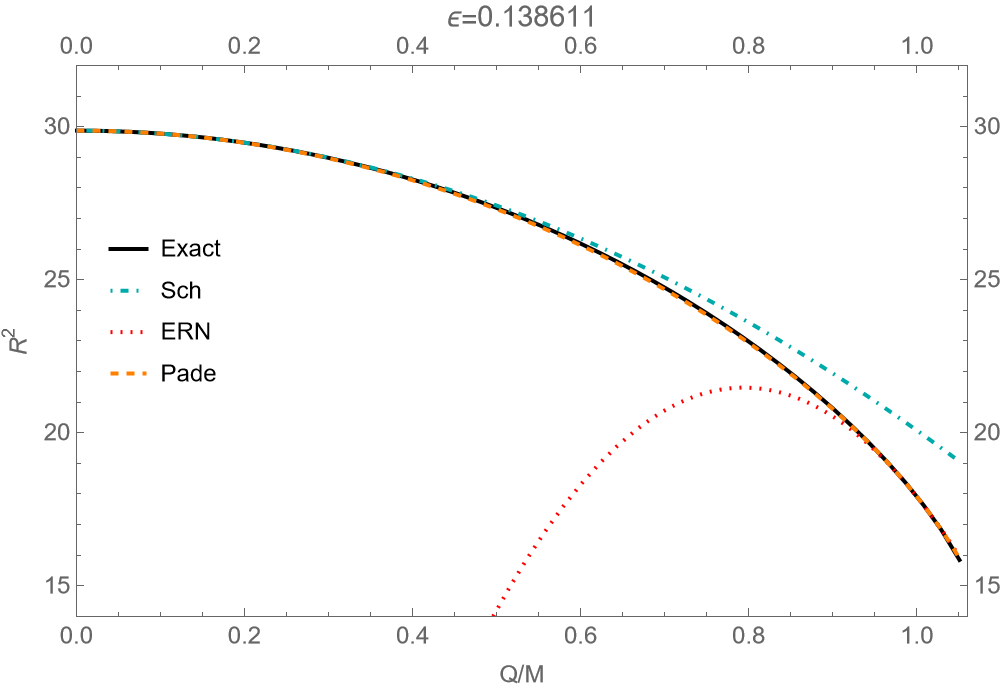}
	\caption{\small{The squared massive shadow radius $R^2$  of the RN black hole as a function of $Q/M$ for $\epsilon=0.138611$. The solid line represents the exact solution, while the dashed line corresponds to the Padé approximant. The dotdashed and dotted line denote the $\delta$-expansions over Schwarzschild metric and the ERN metric, respectively.}}\label{fig2}
\end{figure}

\begin{figure}[ht]
	\centering
	\includegraphics[width=0.8\textwidth]{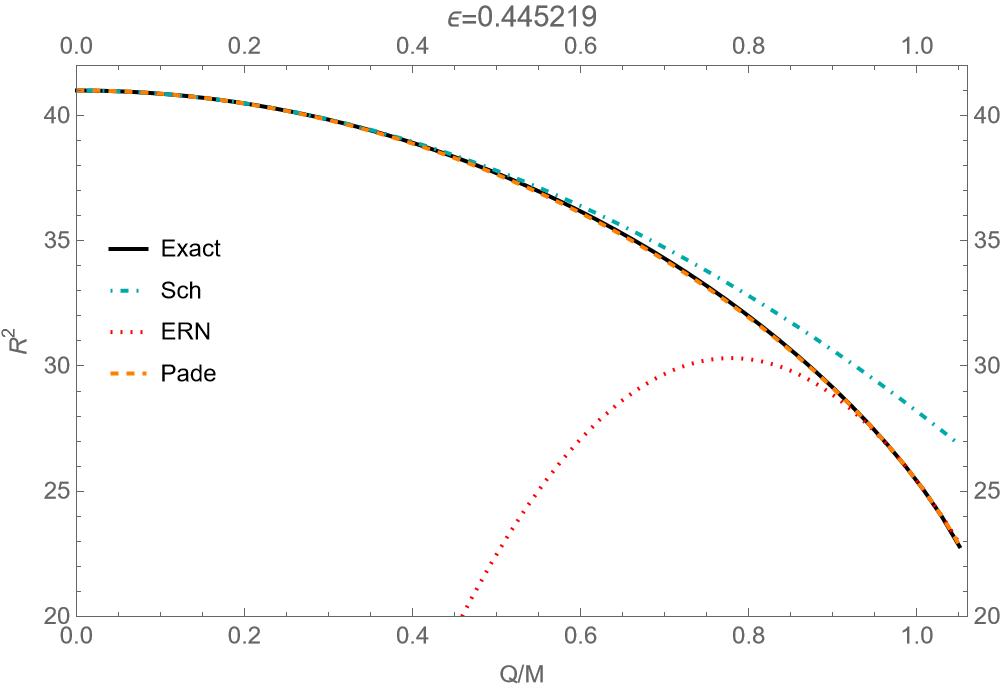}
	\caption{\small{The squared massive shadow radius $R^2$ of the RN black hole as a function of $Q/M$ for $\epsilon=0.445219$. The solid line represents the exact solution, while the dashed line corresponds to the Padé approximant. The dotdashed and dotted line denote the $\delta$-expansions over Schwarzschild metric and the ERN metric, respectively.}}\label{fig3}
\end{figure}

When the expansions in Eqs. (\ref{expan0}) and (\ref{expan1}) are truncated to second order, the continued fraction in Eq. (\ref{padeMRN}) is effectively truncated at $n = 4$ (i.e., $a_n = 0$ for $n > 4$).
By comparing the expansions in (\ref{expan0}) and (\ref{padeMRN}) at \(x=0\), we find that
\begin{eqnarray}
	\label{al0}
	&&A_0 =a_1+B_0-B_1,
	\\\label{al1}
	&&A_1=B_1-a_1\left(a_2+2\right),
	\\\label{al2}
	&&A_2=a_1 \left(a_2^2+a_3 a_2+2 a_2+1\right).
\end{eqnarray}
Similarly, by comparing the expansions in (\ref{expan1}) and (\ref{padeMRN}) at \(x=1\), we obtain
\begin{equation}
	\label{be2}
	B_2=\dfrac{a_1}
	{1+\dfrac{a_2}	{1+\dfrac{a_3}{1+a_4}}}.
\end{equation}
By solving Eqs. (\ref{al0})–(\ref{be2}), we obtain the first four coefficients in Eq. (\ref{padeMRN}) as
\begin{eqnarray}
	\label{a1}
	&&a_1=A_0-B_0+B_1	,
	\\\label{a2}
	&&	a_2=\dfrac{-2 A_0-A_1+2 B_0-B_1}{A_0-B_0+B_1}	,
	\\\label{a3}
	&&	a_3=
	\dfrac{-2 A_0 B_0-2 A_1 B_0+A_2 B_0-A_2 B_1+A_0^2+2 A_1 A_0-A_2
		A_0+A_1^2+B_0^2}{\left(2 A_0+A_1-2
		B_0+B_1\right) \left(A_0-B_0+B_1\right)}	,
	\\\label{a4}
	&&	a_4=-\dfrac{\left(A_0-B_0+B_1\right) }
	{\left(2 A_0+A_1-2
		B_0+B_1\right) 
		\left[\left(A_0-B_0+B_1\right){}^2+B_2 \left(A_0+A_1-B_0\right)\right]}
\\\nonumber
&&\quad\times\left[A_0\left(3 A_1-A_2-6 B_0+3 B_1+B_2\right)+A_2 B_0-A_2 B_1+B_2
	\left(A_2-B_0\right)+A_1 \left(-3 B_0+B_1+B_2\right)
\right.
\\\nonumber
&&\qquad\left.
+3 A_0^2+A_1^2+3 B_0^2+B_1^2-3 B_0 B_1\right].
\end{eqnarray}
Therefore, we obtain the $n = 4$  Padé approximant for
the squared massive shadow radius of the RN black hole.
Figs. \ref{fig1}–\ref{fig3} show the squared gravitational shadow radius \(R^2\) of the RN black hole
as a function of \(Q/M\) for various values of the energy parameter \(\epsilon\).
As shown in the three figures, noticeable deviations from the exact solution are found for the dot‑dashed line (the \(\delta\)-expansion about the Schwarzschild metric) when \(Q\) is away from \(0\), and for the dotted line (the \(\delta\)-expansion about the ERN metric) when \(Q\) differs from \(M\).
In contrast, the dashed line, which denotes the Padé approximant, agrees remarkably well with the exact solution.
This demonstrates that the Padé approximant provides an excellent approximation to the exact gravitational shadow radius.

Then we apply the method from Ref.\cite{Kobialko:2024zhc} to reconstruct the parameter \(x~(Q/M)\) of the RN black hole.
To determine \(x~(Q/M)\), we employ  $n = 4$  Padé approximant
\begin{equation}
	\label{padeMRN1}
R^2_{\text{Padé}}(x, \epsilon)=B_0+B_1 (x-1)
	+\dfrac{a_1}
	{1+\dfrac{a_2 x}
		{1+\dfrac{a_3 x}{1+a_4 x}
		}
	} (x-1)^2.
\end{equation}
Assuming that several measurements of the gravitational shadow radii \(R_j\) are available for a set of particles with different energy parameters \(\epsilon_j\).
Then we can obtain a set of approximate relations
\begin{equation}\label{Rpj}
R^2_{\text{Padé}}(x, \epsilon_j)
=
R^2_{j},\quad
j=0,1,\cdots,n,
\end{equation}
where the index \(j=0\) corresponds to the photon shadow case, with $\epsilon_0=0$.
One can then construct the dimensionless ratio
\begin{equation}\label{chi}
\chi_j=\dfrac{R^2_{j}}{R^2_{0}},\quad
j=1,\cdots,n,
\end{equation}
which directly cancels out the dependence on the black hole distance in the observable.
From Eqs. (\ref{Rpj}) and (\ref{chi}), we obtain
\begin{equation}\label{Rpj1}
\dfrac{R^2_{\text{Padé}}(x, \epsilon_j)}
{R^2_{\text{Padé}}(x, 0)}
=
\chi_{j},\quad
	j=1,\cdots,n.
\end{equation}
Theoretically, one can determine several unknown parameters by using a sufficiently large set of measurements \(R_j\) and solving the system of equations (\ref{Rpj1}).
In the present case, only one parameter is involved, and hence a single equation is adequate to fix it.
Admittedly, the determination of the value \(Q\) necessarily relies on the black hole mass \(M\).

Subsequently, we use the exact formula (\ref{R}) to generate simulated ``experimental'' samples for six different values of \(Q/M\), namely, 0.1, 0.3, 0.5, 0.7, 0.8, and 0.9.
For ease of comparison, we set the values of the energy parameter, which are the same as those in Ref.\cite{Kobialko:2024zhc}, to be \(\epsilon_{\text{exp}1}=0.138611\) and \(\epsilon_{\text{exp}2}=0.445219\), respectively.
From the resulting gravitational shadow data, we then compute the corresponding \(\chi_{\text{exp}}\).
Substituting these data into Eq. (\ref{Rpj1}) and solving it numerically, we obtain the reconstructed parameter \(Q_{\text{exp}}/M\).
Moreover, we define the relative error associated with the reconstructed parameter, taking the form
\begin{equation}
\Delta=\dfrac{|Q_{\text{exp}}/M-Q/M|}
{Q/M}\times 100\%.
\end{equation}
The reconstructed values of $Q/M$ obtained using the Padé approximant and KG's approach are listed in Table \ref{table2}.
It is found that, based on our Padé approximant, the overall relative errors of the reconstructed parameters are almost all less than 2\%  for \(Q/M\) ranging globally from 0.1 to 0.9, indicating that the approximant achieves a high degree of accuracy.
Although KG's method exhibits good accuracy for small \(Q/M\), its relative errors increase significantly when \(Q/M > 0.7\), reaching about 8\% at \(Q/M = 0.9\). 
Therefore, this method is not ideal for global parameter reconstruction. 
In contrast, our Padé approximant provides sufficient accuracy across the entire parameter range to approximate the true values, 
and its accuracy can be further enhanced by raising its order.

\begin{table*}[ht]
	\centering
\begin{tabular}{|c|c|c|cc|cc|}
\hline
&&&\multicolumn{2}{|c|}{\text{Padé approximant}} 
&\multicolumn{2}{|c|}{\text{KG's approach}}
\\
\cline{4-7}
$Q/M $ &$\epsilon_{\text{exp}}$&
$\chi_\text{exp}$  &$Q_{\text{exp}}/M$     & 
$\Delta$($\%$)&$Q_{\text{exp}}/M$ & $\Delta$($\%$)
\\
\hline
    0.1 & 0.138611 &1.10646 & 0.099278&0.722 &0.100055&0.055\\
        & 0.445219 &1.51894 & 0.099340&0.660 &0.100052&0.052\\
		\hline
	0.3 & 0.138611 &1.10697 & 0.294804 &1.732  &0.301549&0.516\\
		& 0.445219 &1.52160 & 0.295256 &1.581  &0.301472&0.491\\
		\hline
	0.5 & 0.138611 &1.10811 & 0.489620 &2.076  &0.507879&1.576\\
		& 0.445219 &1.52756 & 0.490559 &1.888 &0.507503&1.501\\
		\hline
	0.7 & 0.138611&1.11027  & 0.688958   &1.577 &0.725448&3.635\\
		& 0.445219 &1.53879 & 0.690040   &1.423 &0.724313&3.473\\
		\hline
	0.8 & 0.138611&1.11204  & 0.792172   &0.979 &0.842926&5.366\\
		& 0.445219&1.54798  & 0.792991   &0.876 &0.841087&5.136\\
		\hline
	0.9 & 0.138611&1.11477  & 0.897380   &0.291 &0.972534&8.059\\
		& 0.445219&1.56203  & 0.897676   &0.258 &0.969531&7.726\\
		\hline
	\end{tabular}
\caption{Reconstruction of $Q/M$ by Padé approximant and Kobialko's approach.}\label{table2}
\end{table*}

%-----------------

%%%%%%%%%%%%%%%%%%%%%%%%%%%%%%%%%%%%%%%%%%%%%%%%%%%%%%%%%%%%%%%%%%%%%%%%%%%%
\section{Conclusions}\label{sec:conclu}

In this paper, we rederive the perturbation theory of gravitational shadows for static spherically symmetric spacetimes by introducing two auxiliary functions \(F(r,\epsilon,\delta)\) and \(G(r,\epsilon,\delta)\).
Our formulae can be easily generalized to higher-order terms.
For the first time, we adopt the ERN black hole instead of the Schwarzschild black hole as the background for perturbative expansions, and systematically derive the two-parameter perturbative expansions of the MPS radius and the squared massive shadow radius.
Then, we apply this method to four spacetime geometries beyond general relativity, and derive the expansion coefficients about the ERN metric.
Notably, even if all models share the same \(\epsilon\) dependence, their differences in \(\delta\) related coefficients still provide observable discriminant features, highlighting the necessity of multi‑parameter joint measurements.
Furthermore, a scanning window is closer to stronger gravitational regimes, because the ERN photon sphere radius \(2M\) is significantly smaller than the Schwarzschild value of \(3M\), so the radial scanning range achieved by varying \(\epsilon\) covers stronger gravitational regions, making it more sensitive to spacetime geometry details.

The asymptotic expansions of the massive shadow radius derived from the Schwarzschild and ERN backgrounds exhibit notable deviations from the exact solution when the charge parameter \(Q\) deviates from the respective expansion points. 
To overcome this limitation, a two‑point Padé approximant is constructed in the form of a continued fraction, which globally matches the expansions at both endpoints. 
This approximant is found to be in excellent agreement with the exact solution across the entire range, and is superior to single perturbative expansions.
We apply the Padé approximant to reconstruct the RN parameter \(Q/M\), using the ratio of shadow radii to remove distance dependence. 
Our mock tests show that the Padé reconstruction maintains relative errors almost all below 2\% throughout the parameter range, whereas existing methods give errors surging to about 8\% for \(Q/M > 0.7\).
In summary, the two‑point Padé approximant not only serves as a highly accurate global approximation for the massive shadow radius, but also provides a reliable basis for model‑independent metric parameter reconstruction.
	
Based on the results of this work, we identify the following directions for further investigation.
The present static spherically symmetric framework should be generalized to axisymmetric spacetimes, such as Kerr or Kerr-Newman black holes and their extremal cases, to study the effects of rotation on the perturbative expansion of massive shadows.
Our perturbative formulae can be applied to actual multi‑frequency EHT data as the ngEHT and other related projects progress. 
The goal is to perform spectroscopic analysis of the shadows of Sgr A* and M87* and to extract model‑independent constraints on metric parameters.

%%%%%%%%%%%%%%%%%%%%%%%%%%%%%%%%%%%%%%%%%%%%%%%%%%%%%%%%%%%%%%%%%%%%%%%%%%%%
\begin{acknowledgments}
This work is supported by the innovation program of Shanghai Normal University under Grant No.~KF202472.
\end{acknowledgments}
%%%%%%%%%%%%%%%%%%%%%%%%%%%%%%%%%%%%%%%%%%%%%%%%%%%%%%%%%%%%%%%%%%%%%%%%%%%%
\appendix
\section{Expressions for the Second-Order Partial Derivatives of the MPS Radius}\label{sec:app}

This appendix provides the complete analytic expressions for the second-order partial derivatives of the MPS radius \(r = r(\epsilon,\delta)\) defined by the auxiliary function \(F(r,\epsilon,\delta)\), with respect to the particle energy parameter \(\epsilon\) and the metric deviation parameter \(\delta\). These formulae were derived in Section \ref{sec:massive} by the implicit differentiation rule and constitute the core computational tools for the perturbative expansion of massive shadows. Substituting Eq. (\ref{F}) into Eqs. (\ref{ddr1})-(\ref{ddr3})  and simplifying yields the second-order partial derivatives of \(r(\epsilon,\delta)\) as follows:

\begin{eqnarray}
	\nonumber
	&&{r_{,\epsilon\epsilon}=}
	-\alpha ^5 \left\{-6 \beta  \alpha _{,r}^3 \beta _{,r}^2+\alpha _{,r}^2 \left(6
	\alpha  \beta _{,r}^3-4 \alpha  \beta  \beta _{,r} \beta
	_{{,rr}}\right)
	-\alpha ^2 \beta _{,r} \left[\alpha _{{,rr}} \left(3
	\beta _{,r}^2-2 \beta  \beta _{{,rr}}\right)+\beta  \beta _{,r} \alpha
	_{{,rrr}}\right]
	+6\alpha  \alpha _{,r} \alpha_{{,rr}}
	\beta  \beta _{,r}^2 
	%--------
	\right.
	\\\label{ree}
	&&\quad
	\left.	
	+\alpha  \alpha _{,r} \left[
	+\alpha  \beta _{{,rr}} \left(\beta _{,r}^2-2 \beta  \beta
	_{{,rr}}\right)+\alpha  \beta  \beta _{,r} \beta
	_{{,rrr}}\right]
	\right\}
	%--------
	\bigg/
	%--------
	\left\{
	\left[2 \beta  \alpha _{,r}^2+\alpha _{,r}
	\left(\dfrac{\alpha  \beta  \beta _{{,rr}}}{\beta _{,r}}-2 \alpha  \beta
	_{,r}\right)-\alpha  \beta  \alpha _{{,rr}}\right]^3
	\right\},
	\\\nonumber
	&&{r_{,\delta\delta}=}
	\left\{
	\left[\alpha  \beta  \alpha _{,r} \beta _{{,r\delta}}-\beta _{,r}
	\left(\alpha  \alpha _{,\delta} \beta _{,r}+\alpha _{,r} \left(\alpha  \beta
	_{,\delta}-2 \beta  \alpha _{,\delta}\right)+\alpha  \beta  \alpha
	_{{,r\delta}}\right)\right]^2 
	%-----
	\left[-6 \beta  \alpha_{,r}^3 \beta_{,r}^2
	+\alpha_{,r}^2 \left(6 \alpha  \beta _{,r}^3-4 \alpha \beta\beta _{,r}
	\beta_{{,rr}}\right)\right.\right.
	%-----
	\\\nonumber
	&&\qquad\left.
	-\alpha^2 \beta _{,r} \left(\alpha _{{,rr}}
	\left(3 \beta _{,r}^2-2 \beta  \beta _{{,rr}}\right)
	+\beta  \beta _{,r}\alpha _{{,rrr}}\right)
	+\alpha \alpha_{,r} \left(6\beta\beta _{,r}^2 \alpha _{{,rr}}
	+\alpha  \beta _{{,rr}} \left(\beta _{,r}^2-2 \beta 
	\beta _{{,rr}}\right)+\alpha  \beta  \beta _{,r} \beta
	_{{,rrr}}\right)\right]
	%-----
	\\\nonumber
	&&\quad
	-2 \left[\alpha  \beta  \alpha _{,r} \beta _{{,rr}}-\beta _{,r} \left(-2 \beta
	\alpha _{,r}^2+2 \alpha  \alpha _{,r} \beta _{,r}+\alpha  \beta  \alpha
	_{{,rr}}\right)\right]
	%-----
	\left[\alpha  \beta  \alpha _{,r} \beta
	_{{,r\delta}}-\beta _{,r} \left(\alpha  \alpha _{,\delta} \beta
	_{,r}+\alpha _{,r} \left(\alpha  \beta _{,\delta}-2 \beta  \alpha _{,\delta}\right)+\alpha  \beta  \alpha _{{,r\delta}}\right)\right]
	\\\nonumber
	&&\qquad
	\left[-2 \alpha _{,r}^2 \beta _{,r} \left(3 \beta  \alpha _{,\delta} \beta
	_{,r}-\alpha  \beta _{,\delta} \beta _{,r}+\alpha  \beta  \beta
	_{{,r\delta}}\right)+\alpha  \beta _{,r} \left(\alpha _{{,rr}}
	\left(2 \beta  \alpha _{,\delta} \beta _{,r}-\alpha  \beta _{,\delta} \beta_{,r}
	+\alpha  \beta  \beta _{{,r\delta}}\right)+\alpha  \alpha_{{,r\delta}} 
	\left(\beta  \beta _{{,rr}}-2 \beta_{,r}^2\right)\right)
	\right.
	%-----
	\\\nonumber
	&&\qquad\left.
	~+\alpha  \alpha _{,r} \left(4 \beta  \beta _{,r}^2 \alpha
	_{{,r\delta}}+\alpha _{,\delta} \left(4 \beta _{,r}^3-2 \beta  \beta
	_{,r} \beta _{{,rr}}\right)+\alpha  \beta _{,r} \left(\beta _{,\delta}
	\beta _{{,rr}}+\beta  \beta _{{rr\delta}}\right)-2 \alpha 
	\beta  \beta _{{,r\delta}} \beta _{{,rr}}\right)-\alpha ^2
	\beta  \beta _{,r}^2 \alpha _{{,rr\delta}}\right]
	\\\nonumber
	&&\quad
	+\left[\beta _{,r} \left(-2 \beta  \alpha _{,r}^2+2 \alpha  \alpha _{,r} \beta
	_{,r}+\alpha  \beta  \alpha _{{,rr}}\right)-\alpha  \beta  \alpha _{,r}
	\beta _{{,rr}}\right]^2 
	%-------
	\left[-\alpha  \beta _{,r}^2 \left(\alpha 
	\beta  \alpha_{,r\delta\delta}-2 \alpha _{,\delta}^2 \beta _{,r}+\alpha 
	\alpha _{,\delta \delta } \beta _{,r}\right)
	\right.
	\\\nonumber
	&&\qquad\left.
	+\alpha _{,r} \left(
	\alpha 
	\left(\alpha  \left(\beta  \beta _{,r} \beta_{,r\delta\delta}-\beta
	_{,\delta \delta } \beta _{,r}^2+2 \beta _{,\delta} \beta _{,r} \beta
	_{{,r\delta}}-2 \beta  \beta _{{,r\delta}}^2\right)+2 \beta
	\alpha _{,\delta \delta } \beta _{,r}^2\right)
	-6 \beta  \alpha _{,\delta}^2
	\beta _{,r}^2+4 \alpha  \alpha _{,\delta} \beta _{,r} \left(\beta _{,\delta}
	\beta _{,r}-\beta  \beta _{{,r\delta}}\right)
	\right)
	\right.
	\\\label{rdd}
	&&\qquad~\left.\left.
	+2 \alpha  \beta
	_{,r} \alpha _{{,r\delta}} \left(2 \beta  \alpha _{,\delta} \beta
	_{,r}-\alpha  \beta _{,\delta} \beta _{,r}+\alpha  \beta  \beta
	_{{,r\delta}}\right)\right]
	\right\}
	\bigg/
	\left\{
	\alpha\beta_{,r}^7 
	\left[
	2 \alpha\alpha_{,r}
	-\frac{\beta\left(\beta_{,r} \left(2 \alpha_{,r}^2-\alpha  \alpha _{{,rr}}\right)+\alpha\alpha_{,r} \beta_{{,rr}}\right)}
	{\beta _{,r}^2}
	\right]^3
	\right\},
	\\\nonumber
	&&{r_{,\epsilon\delta}=}
	\left\{
	\alpha ^2 \left[\alpha  \beta  \alpha _{,r} \beta _{{,r\delta}}-\beta _{,r}
	\left(\alpha  \alpha _{,\delta} \beta _{,r}+\alpha _{,r} \left(\alpha  \beta
	_{,\delta}-2 \beta  \alpha _{,\delta}\right)+\alpha  \beta  \alpha
	_{{,r\delta}}\right)\right]
	\left[-6 \beta  \alpha _{,r}^3 \beta
	_{,r}^2+\alpha _{,r}^2 \left(6 \alpha  \beta _{,r}^3-4 \alpha  \beta  \beta _{,r}\beta _{{,rr}}\right)
	%-----
	\right.\right.
	\\\nonumber
	&&\qquad
	\left.
	-\alpha ^2 \beta _{,r} \left(\alpha _{{,rr}}
	\left(3 \beta _{,r}^2-2 \beta  \beta _{{,rr}}\right)+\beta  \beta _{,r}
	\alpha _{{,rrr}}\right)+\alpha  \alpha _{,r} \left(6 \beta  \beta _{,r}^2
	\alpha _{{,rr}}+\alpha  \beta _{{,rr}} \left(\beta _{,r}^2-2 \beta 
	\beta _{{,rr}}\right)+\alpha  \beta  \beta _{,r} \beta
	_{{,rrr}}\right)\right]
	%----------
	\\\nonumber
	&&\quad
	-\alpha ^2 \left[\alpha  \beta  \alpha _{,r} \beta _{{,rr}}-\beta _{,r} \left(-2 \beta 
	\alpha _{,r}^2+2 \alpha  \alpha _{,r} \beta _{,r}+\alpha  \beta  \alpha
	_{{,rr}}\right)\right]
	%--------
	\\\nonumber
	&&\qquad
	\left[
	\alpha  \beta _{,r} 
	\left(\alpha_{{,rr}} \left(2 \beta  \alpha _{,\delta} \beta _{,r}-\alpha  \beta
	_{,\delta} \beta _{,r}+\alpha  \beta  \beta _{{,r\delta}}\right)+\alpha  \alpha _{{,r\delta}} \left(\beta  \beta
	_{{,rr}}-2 \beta _{,r}^2\right)\right)
	\right.
	%--------
	\\\nonumber
	&&\qquad
	\left.
	+\alpha  \alpha _{,r} \left(4 \beta 
	\beta _{,r}^2 \alpha _{{,r\delta}}+\alpha _{,\delta} \left(4 \beta
	_{,r}^3-2 \beta  \beta _{,r} \beta _{{,rr}}\right)
	+\alpha \beta _{,r}\left(\beta _{,\delta} \beta _{{,rr}}+\beta  \beta _{{,rr\delta}}\right)-2 \alpha  \beta  \beta _{{,r\delta}} \beta
	_{{,rr}}\right)-\alpha ^2 \beta  \beta _{,r}^2 \alpha _{{,rr\delta}}
	\right.
	\\\label{red}
	&&\qquad
	\left.\left.
	%----
	-2 \alpha _{,r}^2 \beta _{,r} 
	\left(3 \beta 
	\alpha _{,\delta} \beta _{,r}-\alpha  \beta _{,\delta} \beta _{,r}+\alpha 
	\beta  \beta _{{,r\delta}}\right)	
	\right]\right\}
	\bigg/
	\left\{
	\beta _{,r}^5 \left[\frac{\beta  \left(\beta _{,r} \left(2 \alpha _{,r}^2-\alpha 
		\alpha _{{,rr}}\right)+\alpha  \alpha _{,r} \beta
		_{{,rr}}\right)}{\beta _{,r}^2}-2 \alpha  \alpha _{,r}\right]^3
	\right\}.
\end{eqnarray}

%%%%%%%%%%%%%%%%%%%%%%%%%%%%%%%%%%%%%%%%%%%%%%%%%%%%%%%%%%%%%%%%%%%%%%%%%%%%
\bibliography{PTref}
%%%%%%%%%%%%%%%%%%%%%%%%%%%%%%%%%%%%%%%%%%%%%%%%%%%%%%%%%%%%%%%%%%%%%%%%%%%%
\end{document}